\documentclass[aps,prl,amsmath,twocolumn,amssymb,titlepage,superscriptaddress,showpacs]{revtex4-1}	
\usepackage[english]{babel}
\usepackage{graphicx}
\usepackage{amsmath}
\usepackage{amssymb}
\usepackage{epstopdf}
\usepackage{amsfonts}
\usepackage{bm}
\usepackage{mathdots}
\usepackage{times}
\usepackage{mathtools}
\usepackage{hyperref}

\usepackage{color}

\newcommand{\astcycl}{\mathrlap{\kern0.085em{\circlearrowright}}\ast}
\newcommand{\taucycl}{\mathrlap{\kern0.42em{\bullet}}\circlearrowright}

\begin{document}
\title{Designing spin and orbital exchange Hamiltonians with ultrashort electric field transients}
\author{Martin Eckstein}
\affiliation{Max Planck Institute for the structure and Dynamics of Matter 22761 Hamburg, Germany}
\author{Johan~H.~Mentink}
\affiliation{Radboud University Nijmegen, Institute of Molecules and Materials,
Heyendaalseweg 135, 6525 AJ Nijmegen, The Netherlands}
\author{Philipp Werner}
\affiliation{Department of Physics, University of Fribourg, 1700 Fribourg, Switzerland}


\begin{abstract}
We demonstrate how electric fields with arbitrary time profile can be used to control the time-dependent parameters of spin and orbital exchange Hamiltonians. Analytic expressions for the exchange constants are derived from a time-dependent Schrieffer-Wolff transformation, and the validity of the resulting effective Hamiltonian is verified for the case of a quarter-filled two-orbital Hubbard model, by comparing to the results of a full nonequilibrium dynamical mean-field theory simulation. The ability to manipulate Hamiltonians with arbitrary time-dependent fields, beyond the paradigm of Floquet engineering, opens the possibility to control intertwined spin and orbital order using laser or THz pulses which are tailored to minimize electronic excitations.
\end{abstract}
\maketitle

Femtosecond laser pulses provide intriguing opportunities for manipulating and even switching between phases of complex materials on ultrafast time scales \cite{giannetti2016,kirilyuk2010}. Many successful scenarios to trigger ultrafast phase transitions rely on the excitation of non-equilibrium electron distributions, i.e. by photo doping or ultrafast heating \cite{ostler2012,stojchevska2014,Wegkamp2014,mor2016,stupakiewicz2017}. As a result, the photo-induced dynamics proceeds by relaxation of photo-excited carriers that is irreversible on ultrafast time scales. Clearly, a highly appealing scenario beyond photo-excitation is to achieve full control of the quantum many-body dynamics even during the excitation pulse, which could enable a reversible manipulation of optically induced phase transitions.

In recent years, the problem of such ultrafast and reversible control of solids has been attacked by engineering light-dressed Hamiltonians with off-resonant time-periodic perturbations. The evolution of a system over one period of the external field is described by a  so-called Floquet Hamiltonian $H_F$, which can be manipulated by the amplitude and frequency of the field.  This can be used to control tunneling \cite{Dunlap1986}, change the topology of bands \cite{Oka2009,Kitagawa2011,Wang2013,Jotzu2015}, and to manipulate many-body interactions such as spin-exchange \cite{Mentink2015,Mikhaylovskiy2015,Bukov2016} or superconducting pairing \cite{Knap2016,Komnik2016,Coulthard2016,Murakami2017}.
%
In this letter, we would like to extend on this in two directions: First, (i), we investigate the potential of designing light-induced interactions in the particularly interesting case of intertwined spin and orbital order, which is a hallmark of correlated materials \cite{Tokura2000}. Orbital interactions are often frustrated, which leads to rich phase diagrams, and nontrivial light-induced dynamics, including switching to hidden states \cite{Polli2007,Rini2007,Wall2009,Foerst2011PRB,Beaud2014,Ichikawa2011}. Secondly, (ii), when trying to make use of light-induced interactions in solids, a fundamental limit is set by the energy absorption from the drive. Theoretically, $H_F$ can be derived in the ideally off-resonant limit of high-frequencies \cite{Goldman2014,Bukov2015,Eckardt2015,Itin2015,Mikami2016}, but in real solids absorption is usually low only in small frequency windows. As an alternative, strong few-cycle optical or THz pulses are available for the control of solids \cite{Kampfrath2011,Schubert2014,Ishikawa2014,Baierl2016PRL,Baierl2016}, so it is an interesting question whether light-dressed Hamiltonians can be generalized to transient pulses, down to the singe-cycle limit, and how to optimize the pulse parameters.
%
In this paper, we address these questions by first generalizing the control of spin-exchange Hamiltonians to the case of electric field pulses of arbitrary shape, and apply this approach to the manipulation of Hamiltonians with coupled spin and orbital exchange interactions, i.e. the celebrated Kugel-Khomskii interactions \cite{Kugel1982}. 
%


A standard approach to derive low-energy Hamiltonians are perturbative unitary transformations, which remove the transition matrix elements between high-energy states (charge excitations) and low energy states, as in the derivation of the Heisenberg spin model from the Hubbard model \cite{Harris1967}. One can generalize the approach by searching for a {\em time-dependent unitary transformation} to a rotating frame, in which the Hamiltonian does not mix the charge and spin sector {\em at any time}. In the rotating frame, the dynamics of the low-energy sector is governed by time-dependent exchange interactions that depend on time-dependent electric fields.
To carry out this program, which was already done for an attractive Hubbard model \cite{Sota2016} (and for the time-periodic case \cite{Canovi2016,Bukov2016}),  we separate the Hilbert space into the low-energy and excited sector $\mathcal{H}_0$ and $\mathcal{H}_1$, respectively, with projectors $\mathcal{P}_0$ and $\mathcal{P}_1=1-\mathcal{P}_0$, and decompose each operator $A$ into transitions $A_{ab}\equiv \mathcal{P}_a A \mathcal{P}_b$ between and within the sectors 
\cite{Chao1977}. Furthermore, we assume that the Hamiltonian $H = V_{11}  + \alpha T$ has an interaction $V_{11}$ which acts in $\mathcal{H}_1$, and the remainder $T$ is controlled by a small parameter $\alpha \ll 1$. In strong-coupling perturbation theory, e.g., $T$ is the hopping. For any time-dependent unitary transformation  $e^{S(t)}$ (parametrized by the antihermitian matrix $S$), which transforms the wave function like $|\Psi_{rot}(t)\rangle = e^{S(t)} |\Psi(t)\rangle$, the Hamiltonian in the rotated frame is $H_{rot}(t) = e^{S(t)} [H -i\partial_t ]e^{-S(t)}$. A Taylor ansatz $S=\alpha S_1 + \alpha^2 S_2 + \cdots$ yields the series 
\begin{align}
\nonumber
&H_{rot}(t)
=
 V 
+
\alpha 
\big\{
 T
+
[S_1, V]
+
i\dot S_1
\big\}
+
\alpha ^2
\big\{
[S_2, V]
\\
&
+
i \dot S_2
+
[S_1, T]
+
\frac{1}{2}
[S_1,i\dot S_1 + [S_1, V]]
\big\}
+
\mathcal{O}(\alpha^3).
\label{hrotexpansion}
\end{align}
One can now truncate the expansion of $S$ after a given order $n$, and choose $S_m$ such that $H_{rot} = \sum_{m=0}^{n} \alpha^m H^{(m)} + \mathcal{O}(\alpha^{n+1})$ has no mixing terms for $m\le n$, $H^{(m)}_{01}=H^{(m)}_{10}=0$. At first order, we request that the first bracket in \eqref{hrotexpansion} should have no mixing terms. The resulting differential equation for $S_1$ yields  
\begin{align}
\label{s110}
S_{1,10}(t) &= -\int d\bar t\, G_{V}^R(t,\bar t) T_{10}(\bar t),
\end{align}
where we introduced the Green's function
$G_V^R(t,t') = -i e^{-i [ V-i0^+](t-t')} \theta (t-t')$
($0^+$ is a positive infinitesimal).
When this expression is inserted into the next order, we have
\begin{align}
\label{h002}
H^{(2)}_{00}
&=
-\tfrac12\big(
 T_{01} S_{1,10} 
+ h.c.
 \big).
\end{align}
Equations \eqref{s110} and \eqref{h002} constitute the general expression for any time-dependent low-energy model to second order, which will now be evaluated for the case of spin and orbital exchange interactions.

As a first illustration, we consider the one-band half-filled Hubbard model
\begin{align}
 &H(t)=-t_0\!\sum_{\langle ij \rangle\sigma}(e^{i\phi_{ij}(t)}c_{i\sigma}^{\dagger}c_{j\sigma}+h.c.)+\sum_{i} U n_{i\uparrow}n_{i\downarrow}.
  \label{Eq.:Hubbard-U}
\end{align}
Here $c_{i\sigma}$ denotes the annihilation operators of a fermion with spin $\sigma$ at the lattice site $i$, $U$ the on-site interaction; $t_0e^{i\phi_{ij}(t)}$ is the hopping integral (restricted to nearest neighbour sites), with a time-dependent Peierls phase $\phi_{ij}(t)=\int^t_0 d\bar t\, \vec{E}(\bar t)(\vec{r}_i-\vec{r}_i)$ that captures the effect of an electric field $\vec{E}(t)$. Taking $V$ and $T$ as the interaction and time-dependent hopping term, respectively, we can evaluate Eqs.~\eqref{s110} and \eqref{h002} assuming $U\gg t_0$. At half-filling, the result is the standard spin-1/2 Heisenberg model $H^{(2)}_{00}=\sum_{(i,j)} J_\text{ex}^{ij}(t) \vec{S}_i \vec{S}_j$, with time-dependent exchange interaction along a bond $(i,j)$  $(U^+=U+i0^+)$,
\begin{align}
\label{exchange}
J_\text{ex}^{ij}(t) = 4t_0^2\,\text{Im}\!\! \int_{-\infty}^t\!\!\! 
\!\!d\bar t \,
e^{iU^+(t-\bar t)}
\cos[\phi_{ij}(t)-\phi_{ij}(\bar t)].
\end{align}
The integral generalizes the energy denominator $1/U$ in the time-independent (zero field ) exchange 
$J_\text{ex,eq}=4t_0^2/U$.

\begin{figure}[t]
\includegraphics[width=\columnwidth]{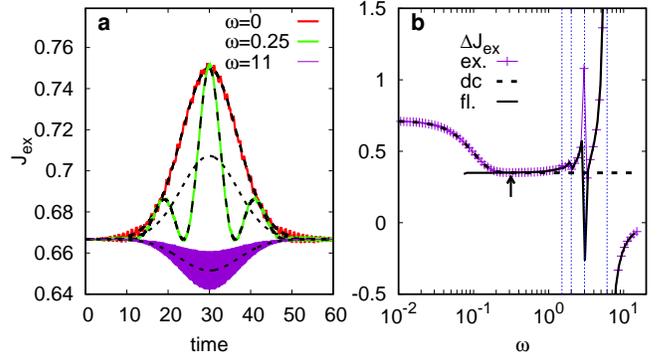}
\caption{(a) Time-dependent exchange interaction [Eq.~\eqref{exchange}]  in the Hubbard model ($U=6$ and $J=1$), for an oscillating electric field $E(t) = E_0 \cos(\omega (t-t_c)) e^{- (t-t_c)^2/t_c^2}$ with gaussian envelope of duration $t_c=10$ and frequency $\omega$. Dashed and dotted lines refer to the instantanous dc and Floquet exchange $J_\text{ex,dc}(E(t))$ and $J_\text{ex,fl}(E_\text{env}(t)/\omega,\omega)$, respectively. (b) Modification of $J_\text{ex}(t)-J_\text{ex}(0)$, integrated over the duration of the pulse. The exact result (symbols) is compared to the instantaneous  dc (dashed line) and Floquet (solid line) expressions. Vertical dotted lines indicate the resonances $n\omega=U$ of the Floquet exchange, the arrow indicates the single-cycle pulse frequency $\omega=\pi/t_c$.}
\label{fig:01}
\end{figure}

In Fig.~\ref{fig:01}a we plot Eq.~\eqref{exchange} for an oscillating field $E(t)$ with gaussian envelope. The time-dependent field generates a $J_\text{ex}(t)$ which has oscillatory and non-oscillatory components, but deviates on average from the equilibrium value $4t_0^2/U$. One easily verifies that Eq.~\eqref{exchange} reproduces the  limits of time-independent and time-periodic fields: A dc electric field with projection $E_{ij}$ along the bond $(i,j)$, $\phi_{ij}(t) = -E_{ij}t$, gives $J_\text{ex,dc}(E_{ij})=2t_0^2/(U-E_{ij})+2t_0^2/(U+E_{ij})$. This dc modification of exchange has been derived from time-independent perturbation theory (taking into account an electrostatic potential), and was observed in cold atom experiments \cite{Duan2003}. Furthermore, for a time-periodic field 
$E_{ij}(t)=A_{ij}\omega\cos(\omega t)$ ($\phi_{ij}(t)=A_{ij}\sin(\omega t)$), Eq.~\eqref{exchange} yields, up to terms oscillating with frequency $\omega$, the exchange interaction $J_\text{ex,fl}(A_{ij},\omega)= \sum_{n} \frac{4t_0^2\mathcal{J}_{|n|}(A_{ij})^2}{U-n\omega}$ known from Floquet theory \cite{Mentink2015} ($\mathcal{J}_n(x)$ is the Bessel function).

Using the general expression \eqref{exchange} it is interesting to check the range of validity of the Floquet expression in the experimentally relevant regime of few-cycle pulses. For this we compare the exact expression to the dc result $J_\text{ex,dc}(E(t))$, evaluated using the instantaneous electric fields, as well as to the Floquet result $J_\text{ex,fl}(E_\text{env}(t)/\omega,\omega)$, evaluated at the instantaneous envelope of the field (see dashed and dotted lines in Fig.~\ref{fig:01}a). To quantify the difference between the expressions, we compare the integral $\Delta J_\text{ex} = \int dt (J_\text{ex}(t)-J_\text{ex}(0)) $ in the three cases,  see Fig.~\ref{fig:01}b. This quantity is related to a possible induced spin dynamics during the pulse. Remarkably, the Floquet expression works down to the limit of a single cycle pulse ($\omega\approx\pi/t_c$), except close to the resonances $n\omega=U$ (vertical lines). One can also see that there are various choices of the field amplitude and frequency with the same effect on $J_\text{ex}$, which leaves room for optimizing the pulse parameters in order to minimize the electronic excitations. This will be addressed below for the case of field-induced orbital dynamics.

As second illustration, we show how Eq.~\eqref{exchange} is generalized to systems with intertwined spin and orbital exchange interactions. We consider $3d$ orbitals in a cubic crystal field, with inactive (filled or empty) $t_{2g}$ states, and one electron in the two-fold degenerate $e_g$ manifold. This situation is described by the two-orbital Hubbard model, with rotationally invariant interaction
$V$ and hopping $T$, 
\begin{align}
\nonumber
&V=
U\sum_{ i,l}
n_{ i l\uparrow}  n_{ i l\downarrow} 
+
\sum_{ i,\sigma\sigma',l\neq l'}
(U'-J_H\delta_{\sigma\sigma'})
 n_{ il\sigma}  n_{ il'\sigma'} 
 \\
&
+
J_H
\sum_{ i,l\neq l'}
\big(c_{ il \uparrow}^\dagger c_{ i l\downarrow}^\dagger
c_{ i l'\downarrow}  c_{ i l'\uparrow} 
+
c_{ i l\uparrow}^\dagger c_{ i l'\downarrow}^\dagger
c_{ i l\downarrow}  c_{ i l'\uparrow} 
\big)
\\
\label{eghopping}
 &T(t)=-\sum_{\langle ij \rangle\sigma,l l'}\big(t_{l l'}e^{i\phi_{ij}(t)}c_{i l\sigma}^{\dagger}c_{j l'\sigma}+h.c.\big).
\end{align}
Here $l=0,1$ labels the $d_{3z^2-r^2}$ and $d_{x^2-y^2}$ orbital, respectively, $J_H$ is the Hunds-coupling,  and $U'=U-2J_H$. As the Coulomb interaction $U$ projects out doubly occupied sites, the model reduces to a low-energy Hamiltonian for spins $\vec S^i$ and orbital pseudo-spins $\vec Z^{i}\equiv(Z^{i}_1,Z^i_2,Z^i_3)$, $Z^i_a=\tfrac12\sum_{\sigma ll'}c_{i l\sigma}^{\dagger}\sigma^a_{l l'} c_{i l'\sigma}$ ($\sigma^{a}$ for $a=1,2,3$ are Pauli matrices), which is the  Kugel-Khomskii model.

We follow Ref.~\cite{Oles2000} to parametrize this model \cite{footnote01}. Since the hopping is rotationally invariant in spin, the exchange Hamiltonian on each bond $(i,j)$ can be factorized as
\begin{align}
h^{ij}
=
(\tfrac34+\vec S^{i} \vec S^{j})
K_T^{ij}
+
(\tfrac14-\vec S^{i} \vec S^{j})
K_S^{ij},
\end{align}
where the terms in brackets are projectors on the spin triplet and singlet, respectively, and $K_{S,T}$ is the orbital part of the Hamiltonian. Due to the geometry of the orbitals, the latter depends on the direction of the bond. In a cubic crystal, hopping $t_0$ along the $z$ direction is possible only between the $d_{3z^2-r^2}$ orbitals, so that $K_{S,T}^{z}$ takes a simple form
\begin{align}
K_{S}^{ij,z} &=  J_{S1}^{z}(Z^{i}_3 Z^{j}_3-\tfrac14)-(J_1^{z}+J_{S2}^{z})(\tfrac12 + Z^{i}_3)(\tfrac12 + Z^{j}_3),
\nonumber\\
\label{htz}
K_{T}^{ij,z} &= J_T^{z}(Z^{i}_3 Z^{j}_3-\tfrac14).
\end{align}
Here $J_s^{z}=2t_0^2 / \varepsilon_s$ correspond to the possible energies  $\varepsilon_{S1}=U+J_H$, $\varepsilon_{S2}=U-J_H$, $\varepsilon_T=U-3J_H$ of a doublon, which appear as virtual states in second-order perturbation theory. Along the $x$ and $y$ bonds, $\vec Z$ spinors have to be replaced by  the spinors $\vec X$ and $\vec Y$ corresponding to ($d_{3x^2-r^2},d_{y^2-z^2})$ and $(d_{3y^2-r^2},d_{z^2-x^2})$, which relate to $\vec Z$ by a $120^\circ$ rotation around the orbital pseudo-spin-$2$ axis, $X_3= -\tfrac12 (-\sqrt{3}Z_1+Z_3)$, $Y_3= -\tfrac12 (\sqrt{3}Z_1+Z_3)$. The orbital part of \eqref{htz} is thus highly anisotropic and favors mutual alignment in different orbitals along different directions, which is an intrinsically frustrated so-called Kompass model \cite{Nussinov2015}.

\begin{figure}[t]
\includegraphics[width=\columnwidth]{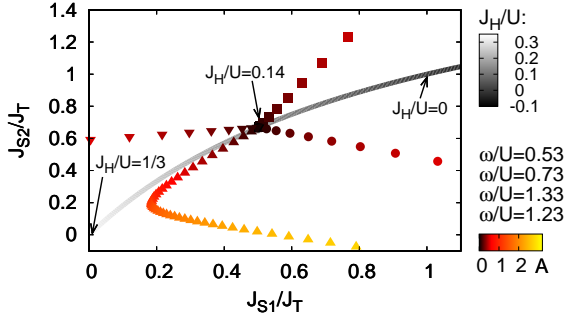}
\caption{Ratio of the (period-averaged) exchange constants $J_{1,2,3}(A,\omega)$ in 
Eq.~\eqref{kkj} for a time periodic field $E(t)=A\omega\cos(\omega t)$ along a given bond, for $J_H/U=0.14$. 
The symbols indicate different frequencies, and the color map shows the amplitude. The line shows the accessible parameter space in equilibrium, for different values of $J_H$ as indicated by the gray-scale. 
}
\label{fig:02}
\end{figure}

In the time-dependent case [Eq.~\eqref{h002}], it is important to note that the time-dependence of the hopping enters as a global (orbital independent) Peierls factor in Eq.~\eqref{eghopping}, and $V$ is time independent. The geometrical structure of the Hamiltonian is therefore unchanged, while the exchange constants $J_{s}^\alpha$ along the $\alpha=x,y,z$ bonds are obtained by replacing the energy denominators $1/\varepsilon_s$ by a time-integral as in Eq.~\eqref{exchange},
\begin{align}
\label{kkj}
J_s^{\alpha}(t)= 2t_0^2\,\text{Im}\!\! \int_{-\infty}^t\!\!\!  \!\!d\bar t \, e^{i\varepsilon_s^+(t-\bar t)} \cos[\phi_{\alpha}(t)-\phi_{\alpha}(\bar t)].
\end{align}

This time-dependent manipulation of the exchange in the Kugel-Khomskii model suggests various  possibilities to act on spin and orbital order. For example, a time-dependent field allows to tune separately the ratios of the three exchange constants, as illustrated in Fig.~\ref{fig:02} for a time-periodic field. While in equilibrium the ratio $J_{S1}/J_T$ and $J_{S2}/J_{T}$ is only a function of $J_H/U$, in nonequilibrium, one can both access parameter regimes which correspond to a different ratio $J_H/U$, and those regimes which are inaccessible in equilibrium. 

Below we verify Eq.~\eqref{kkj} by discussing light-induced orbital dynamics. We will focus on a simple situation where the spin direction is entirely polarized, such that only the triplet term in Eq.~\eqref{htz} contributes, which is the antiferro-orbital $120^\circ$ Kompass model. It is instructive to first look at the classical mean-field dynamics of the orbital model.  Assuming anti-ferro-orbital order with $\langle \vec Z^i(t)\rangle\equiv \vec \tau_{\pm}(t)$ for site $i$ on either of the two sublattices (labelled $\pm$), we obtain mean-field equations of motion $\frac{d}{dt} \vec \tau_\pm(t)  = \vec B_\pm(t) \times \vec \tau_\pm(t)$, with an orbital pseudo-magnetic field that has contributions from each of the two neighbors along the $\alpha=x,y,z$ bonds, $B_{\pm,a}(t)=2\sum_{\alpha,b} J^{\alpha}_T(t) \eta^{\alpha}_{ab}\tau_{\mp,b}(t)$; the geometry of the exchange along the different bonds is captured by the tensor
\begin{align}
\eta^{x(y)}=
\frac{1}{4}
\begin{pmatrix}
3 & 0 &\pm \sqrt{3}
\\
0&0&0
\\
\pm\sqrt{3} & 0 & 1
\end{pmatrix},
\eta^{z}=\begin{pmatrix}
0&0&0
\\
0&0&0
\\
0&0&1
\end{pmatrix}.
\end{align}
One can see that in equilibrium, with $J_T^x=J_T^y=J_T^z\equiv J_\text{ex}$, any configuration in the orbital ($Z_1,Z_3$) plane corresponds to an equilibrium solution, with $\vec B_{\pm}=3J_\text{ex}\vec \tau_\mp$. An external field, polarized along one axis, {\em independently} modifies the exchange along the three bond directions, and thus  induces a precessional dynamics
\cite{footnote_mf}.

To confirm the prediction from the strong-coupling analysis \eqref{kkj}, we now solve the nonequilibrium dynamics of the spin-polarized two-band Hubbard model under the influence of an electric field, using nonequilibrium DMFT \cite{aoki2014rev}. This technique, which maps the lattice model onto a set of coupled single-impurity models, has been described elsewhere, and we defer the details of the implementation to the appendix. The model is solved on the cubic  lattice with a  simplified closed form self-consistency, using a hybridization expansion impurity solver \cite{eckstein2010}. We consider the insulating regime at $U-3J_H=6$ and bandwidth $W=4$ at  initial (inverse) temperature $\beta=30$, which corresponds to antiferro-orbital order along the $z$-direction, with $|\langle \vec Z_3\rangle|\approx0.32$. 

\begin{figure}[tbp]
\includegraphics{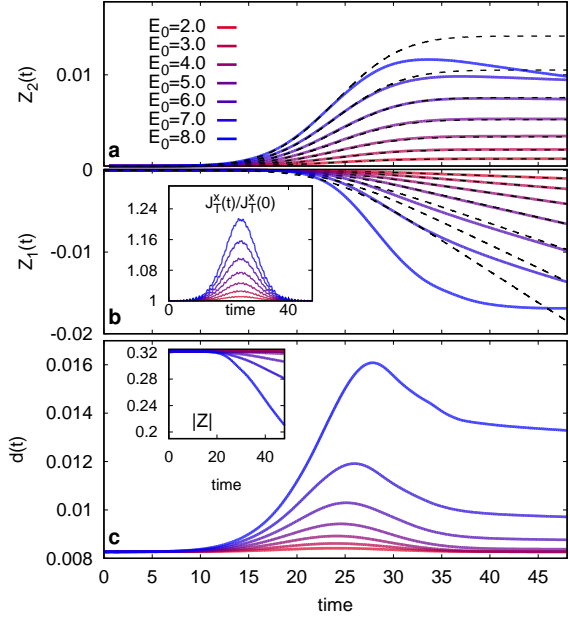}
\caption{
Orbital dynamics induced by an 
electric field $E_x(t)=E_0 e^{- (t-t_c)^2/t_c^2}$ in a cubic environment ($U-3J_H=6$, $J=1$, $\beta=30$). The initial state is anti-ferro orbital order in the orbital pseudo-spin $z$ direction. (a) and (b) $Z_2$ and $Z_1$ for various field amplitudes. The dashed lines correspond to the solution of the mean-field dynamics, with time-dependent exchange interaction given by $\eqref{kkj}$ (see inset in panel b;
$J_T^y(t)=J_T^z(t)=J_T^x(0)$ remain unchanged by the field). (c) Change of the double occupancy, and the total ordered orbital moment $|\vec Z|$ for the same pulses.
}
\label{fig:03}
\end{figure}

Figure \ref{fig:03}a shows the time-evolution of the ordered orbital moment on one sublattice during and after a gaussian field pulse, with polarization along the $x$-direction. For smaller fields ($E_0\le 5$), the evolution of the transverse $Z_{1}$ and $Z_{2}$ components of the orbital order follows the mean-field solution of the spin model, obtained with the time-dependent $J_{T}(t)$ from Eq.~\eqref{kkj} (dashed lines). 

The orbital pseudo-magnetic field $B_{\pm}(t)$ always lies in the $(Z_1,Z_3)$ plane, but due to the bond-dependent exchange coupling (inset in Fig.~\ref{fig:03}b) it is rotated from the original $Z_3$ direction,
such that the precessional motion generates a component $Z_2$, 
which corresponds to a complex superposition of the  $d_{3z^2-r^2}$ and $d_{x^2-y^2}$ orbital. For larger fields, the DMFT results decrease during the pulse compared to the spin model. This can be understood as the strong fields induce charge excitations, which have been projected out in the unitary perturbation theory. This is confirmed in Fig.~\ref{fig:03}c, where we show that the double occupancy $d(t)$ is increased after the pulse (the increase during the pulse corresponds to virtual fluctuations). For a high density of excitations, the orbital ordered moment $|\vec Z|$ starts to decrease after the pulse (see inset), in analogy to the melting of antiferromagnetic spin order by photo-doping \cite{Werner2012}.

\begin{figure}[t]
\includegraphics{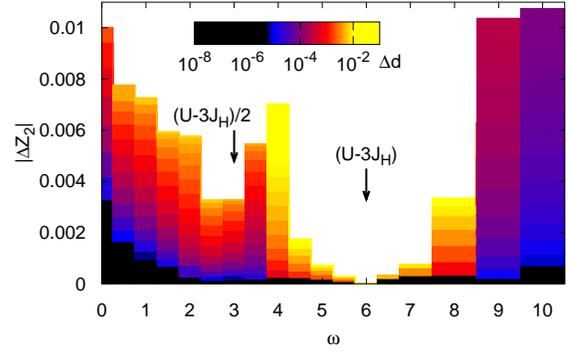}
\caption{For the few-cycle $E(t) = E_0 \cos(\omega (t-t_c)) e^{- (t-t_c)^2/t_c^2}$ with given duration $t_c=8$, the color map shows the excitation density (increase $\Delta d$ of the double occupancy after the pulse), as a function of the frequency, and the achieved orbital precession (change $|\Delta Z_2|$ during the pulse).
}
\label{fig:04}
\end{figure}

Finally, we use the same DMFT setup to find the parameters that realize a desired effect on the ordered state at the least amount of electronic excitation. Figure \ref{fig:04} shows a color map of the absorption, measured in terms of the increase $\Delta d$ of the double occupancy, for pulses within a range of frequencies and field amplitudes, keeping the duration $t_c=8$ of the pulse fixed. The vertical axis is the accumulated precession of the component $Z_2$, which first increases with the field, but cannot exceed a certain maximum as excitations set in at larger amplitudes.  One finds that close to the resonance $U-3J_H=\omega$, and to some extent $U-3J_H=2\omega$, almost no precession can be obtained, as absorption processes strongly compete with the off-resonant dynamics described by the effective exchange Hamiltonian. We observe that to induce a certain effect in the most reversible way, it is best to go to the quasi-dc regime, corresponding to THz excitation in the Mott insulator. Alternatively, one can go to the high-frequency regime, but in real solids there are typically charge excitations to other electronic bands that are not captured in the present simulations.

In conclusion, we have demonstrated the manipulation of spin and orbital exchange Hamiltonians with electric fields of arbitrary time-profile. A time-dependent Schrieffer-Wolff transformation provides an important guiding principle for the evaluation of the effective exchange interactions, as we quantitatively verified by solving the full electronic model which includes the absorption due to field-induced tunnelling or (multi)photon absorption. In particular, in order to control exchange interactions in the most reversible way, our analysis strongly favors low frequency, quasi dc, electric fields corresponding to THz excitation of the Mott insulator. 

These findings suggest many possibilities for future studies. Firstly, orbital ordering is often coupled to Jahn-Teller distortions of the lattice. In this case, the field induced exchange interactions provide one part of the force on the ordering, and it will be interesting to study the combined effect. Moreover, the actual order in the orbital models is not rotationally invariant like in spacial mean field simulations, but entropically stabilized in certain directions due to non-local fluctuations \cite{vdBrinck1999}. Another interesting and open problem is therefore to study the dynamics of a (classical or quantum) model with time-dependent $J_\text{ex}$, taking into account these entropic forces. Finally, our results suggests that it is very interesting to investigate the THz manipulation of exchange Hamiltonians for reversible control of magnetic order at ultralow energy load. 

ME acknowledges support by the DFG within the Sonderforschungsbereich 925 (project B4). PW acknowledges support from FP7 ERC Starting Grant No.~278023. JM was supported by the Nederlandse Organisatie voor Wetenschappelijk Onderzoek (NWO) through a VENI grant.


\section*{Appendix}

In this appendix we give the detailed equations for the implementation of nonequilibrium DMFT for the two-band Hubbard model with  $e_g$ orbitals. We start from the Hamiltonian Eq.~(6) and (7) of the main text, which describes electrons on two orbitals of $e_g$ symmetry at each site, and adopt the parametrization of the orbitals as described in Ref.~\cite{Nussinov2015}, where orbital $1$ and $2$ correspond to orbitals $d_{x^2-y^2}$ and $d_{3z^2-r^2}$, respectively.  We introduce the orbital spinor (omitting site and spin indices for simplicity)
\begin{align}
\hat \psi=
\begin{pmatrix}
c_{1}
\\
c_{2}
\end{pmatrix}
\equiv
\begin{pmatrix}
c_{x^2-y^2}
\\
c_{3z^2-r^2}
\end{pmatrix}.
\end{align}
The Wannier orbitals transform among each other like the atomic $e_g$ orbitals, which have wave functions $\phi_{x^2-y^2}(\vec{r})=\phi(r)(x^2-y^2)$ and $\phi_{3z^2-r^2}(\vec{r})=\phi(r)(3z^2-r^2)/\sqrt{3}$, with some radial part $\phi(r)$. By forming linear combinations of $d_{x^2-y^2}$ and $d_{3z^2-r^2}$, one arrives at the other basis states $d_{y^2-z^2}$, $d_{z^2-x^2}$,  $d_{3x^2-r^2}$, and  $d_{3y^2-r^2}$. These linear combinations can be conveniently represented as rotations in orbital-pseudospin space, generated by ($\sigma_{1,2,3}$ denote the Pauli matrices)
\begin{align}
\hat R(\theta)=e^{i\hat \sigma_2\theta/2 }=
\begin{pmatrix}
\cos(\theta/2) & \sin(\theta/2)
\\
- \sin(\theta/2) & \cos(\theta/2)
\end{pmatrix}.
\end{align}
Rotations by $\theta= 4\pi/3$ give,
\begin{align}
\hat R
\big(
\tfrac43\pi
\big)
\hat \psi
=
\begin{pmatrix}
-\frac{1}{2} c_{x^2-y^2}+\frac{\sqrt{3}}{2}c_{3z^2-r^2}
\\
-\frac{\sqrt{3}}{2} c_{x^2-y^2}-\frac{1}{2}c_{3z^2-r^2}
\end{pmatrix}
=
\begin{pmatrix}
c_{z^2-x^2}
\\
c_{3y^2-r^2}
\end{pmatrix},
\label{rotation1}
\end{align}
where the last equality can be checked by evaluating the operators on the one-particle states, e.g., 
$\langle 0 | -\frac{1}{2} c_{x^2-y^2}+\frac{\sqrt{3}}{2} c_{3z^2-r^2} | \vec{r}\rangle$ 
$= -\frac{1}{2} \phi_{x^2-y^2}(\vec{r})^* +\frac{\sqrt{3}}{2} \phi_{3z^2-r^2}(\vec{r})^*$
$=\phi(r)^* [-\frac{1}{2} (x^2-y^2) +\frac{\sqrt{3}}{2} (2z^2-x^2-y^2)/\sqrt{3}]$
$=\phi_{z^2-x^2}(\vec{r})^* = \langle 0 | c_{z^2-x^2} |\vec{r}\rangle$. In summary, successive application of $\hat R(4 \pi/3)$ corresponds to a permutation of the orbitals $xyz\to zxy \to yzx \to xyz$ \cite{Nussinov2015}.

With the orbital spinor, the hopping is written as
\begin{align}
T=
-
\sum_{ j\sigma}
\sum_{a=x,y,z} 
\Big(
e^{i\phi_a(t)}\,
\hat \psi_{j+\hat e_a,\sigma}^\dagger 
\,\hat v_a\, \hat\psi_{j,\sigma} + h.c.
\Big),
\end{align}
where $\hat e_a$ denotes the unit vector along a bond direction $a$, $\phi_a(t)=\phi_{j+\hat e_a,j}$ is the Peierls phase along the bond ($j+\hat e_a,j$), and $\hat v_a$ is a $2\times2$ matrix. As stated in the main text, we focus on a cubic environment, in which  the only nonvanishing matrix element for hopping along the $z$ axis is between $d_{3z^2-r^2}$ orbitals,
\begin{align}
\hat v_z
=
\begin{pmatrix}
0 & 0
\\
0 & t_0
\end{pmatrix}.
\end{align}
To obtain the hopping along the $x$-bond, we first rotate the lattice around the $y$ axis, which maps $x\to z $ and $z\to -x$, so that the 
hopping along the $x$-bond is then written as
\begin{align}
\begin{pmatrix}
c_{z^2-y^2}^\dagger & c_{3x^2-r^2}^\dagger
\end{pmatrix}
\begin{pmatrix}
0 & 0
\\
0 & t_0
\end{pmatrix}
\begin{pmatrix}
c_{z^2-y^2}
\\ 
c_{3x^2-r^2}
\end{pmatrix}.
\label{vxrotated}
\end{align}
Analogous to Eq.~\eqref{rotation1} we get
\begin{align}
\hat R\big(
-\tfrac43\pi
\big)
\begin{pmatrix}
c_{x^2-y^2}
\\ 
c_{3z^2-r^2}
\end{pmatrix}
=
-\hat \sigma_3
\begin{pmatrix}
c_{z^2-y^2}
\\ 
c_{3x^2-r^2}
\end{pmatrix},
\end{align}
so that Eq.~\eqref{vxrotated} (and the corresponding equation for the $y$-bond) becomes
$\hat \psi^\dagger \hat v_{x/y}\hat \psi$, 
with (the sign $\sigma_3$ cancels)
\begin{align}
\label{hoppx}
\hat v_x
&=
\hat R\big(\tfrac43\pi\big)
 \hat v_z \hat R(-\tfrac43\pi\big)
=
\frac{t_0}{4}
\begin{pmatrix}
3 & -\sqrt{3}
\\
-\sqrt{3} & 1
\end{pmatrix},
\\
\label{hoppy}
\hat v_y
&=
\hat R(-\tfrac43\pi\big) \hat v_z \hat R(\tfrac43\pi\big)
=
\frac{t_0}{4}
\begin{pmatrix}
3 & \sqrt{3}
\\
\sqrt{3} & 1
\end{pmatrix}.
\end{align}

In DMFT, we compute the $2\times2$ contour-ordered Green's function 
\begin{align}
\hat G(t,t')
&=
-i \langle
T_\mathcal{C}
\hat \psi_{\sigma}(t)
\hat \psi_{\sigma}^\dagger(t')
\rangle
\\
&\equiv
\begin{pmatrix}
G_{11}(t,t') & G_{12}(t,t')
\\
G_{21}(t,t') & G_{22}(t,t')
\end{pmatrix}.
\end{align}
(For an introduction into the Keldysh formalism for contour-ordered Green's functions and to nonequilibrium dynamical mean-field theory, see Ref.~\cite{aoki2014rev}.) The DMFT impurity action is given by ($H_{loc}=V$ is the local interaction)
\begin{align}
\label{action}
\mathcal{S}
=
&-i \!\int_{\mathcal{C}} dt\,H_{loc}(t)
-i \sum_\sigma 
\!\int_{\mathcal{C}} \!\!dtdt'\,
\hat\psi^\dagger_\sigma(t)
\hat \Delta(t,t') \hat\psi_\sigma(t'),
\end{align}
with a self-consistently determined hybridization matrix $\hat \Delta$,
and $\hat G$ is determined by
$
\hat G
(t,t')
=-i
\text{tr}
[T_{\mathcal{C}}
e^{\mathcal{S}}
\hat \psi_\sigma(t)
\hat \psi_\sigma^\dagger(t')
]/\mathcal{Z}
$. 

For the DMFT simulation, we focus on a bipartite Bethe lattice in the limit of infinite coordination number $6Z\to\infty $, in which each site has $Z$ bonds attached with rescaled hopping  $-e^{\eta i\phi_a(t)}\hat v_a/\sqrt{6Z}$ for each of the $6$ combinations $\eta=\pm$, $a=x,y,z$. This can be envisioned as the simple limit of an infinitely coordinated lattice with a local cubic environment. As in the single band case, results can be expected to be qualitatively the same as for a three-dimensional cubic lattice with the bandwidth $4t_0$. Denoting by $\hat G_{A,B}$ the Green's function on the two sublattices $A,B$ of the bipartite lattice, the DMFT self-consistency is given by
\begin{align}
\hat \Delta_A(t,t')
&=
\frac{1}{6}\sum_{a=x,y,z}
\Big(
 e^{i\phi_a(t)} \hat v_a\hat G_B(t,t') \hat v_a e^{-i\phi_a(t')}
\nonumber\\
&\,\,\,\,\,\,\,+
e^{-i\phi_a(t)} \hat v_a \hat G_B(t,t') \hat v_a e^{i\phi_a(t')}
\Big)
\label{bethesc}
\end{align}
(the derivation uses the cavity approach, analogous to the single-band case \cite{Georges1996}).

The strong-coupling limit of the $e_g$ Hubbard model is characterized by a entangled spin and orbital dynamics. Here we focus on the case where the spin is fully polarized. We can omit the spin index, and the local Hamiltonian reduces to 
\begin{align}
\label{hlocKK}
H_{loc} = \sum_{j}\big[ (U-3J_H)n_{j, 1}n_{j, 2} -\mu(n_{j, 1}+n_{j, 2})\big].
\end{align}
The model thus becomes equivalent to a spin-less single-band model with orbital-dependent hopping, and we use the standard strong-coupling expansion \cite{eckstein2010} to solve the impurity model in the strongly interacting Mott regime. We allow for Ne\'el-type anti-ferro-orbital sublattice symmetry breaking, and look for solutions in which the Green's function on site $B$ is obtained from site $A$ by a $\pi$ rotation in orbital space,
\begin{align}
\label{NeelAFO}
\hat G_B(t,t') = \hat R(\pi) \hat G_A(t,t') \hat R(-\pi),
\end{align}
which closes the DMFT equations.

It is instructive to verify the symmetries for the DMFT equations \eqref{action}, \eqref{bethesc}, \eqref{hlocKK},  \eqref{NeelAFO}: One can show that in equilibrium (i.e., for $\phi_a=0$) the solution is rotationally invariant around the $\sigma_2$ axis in orbital pseudospin, i.e., if $\hat G$ is a solution, $\hat G(\theta)\equiv \hat R(\theta) \hat G \hat R(-\theta)$ is a solution as well for all $\theta$. To show this we rotate the spinors,
$\hat \psi(\theta)\equiv R(\theta) \hat \psi$, so that $
\hat G(\theta)(t,t')
=-i
\text{tr}
[T_{\mathcal{C}}
e^{\mathcal{S}}
\hat \psi(\theta)(t)
\hat \psi(\theta)^\dagger(t')
]/Z$.
The interaction is rotationally invariant, 
$H_{loc}[\hat\psi^\dagger,\hat\psi]=H_{loc}[\hat\psi^\dagger(\theta),\hat\psi(\theta)]$, and the hybridization can be written as
$-i \int_{\mathcal{C}} dtdt'
\hat\psi^\dagger(\theta)(t)
\hat R(\theta)
\hat \Delta(t,t')\hat R(-\theta) 
\hat\psi(\theta)(t')$. We can see that $\hat G(\theta)(t,t')$ is a solution of the DMFT equations if the rotated hybridiyation function 
$\hat R(\theta)
\hat \Delta_A(t,t')
\hat R(-\theta) 
$ satisfies the self-consistency \eqref{bethesc} with the rotated Green's function, i.e.,
\begin{align}
\hat R(\theta) \Big[
\sum_{a=x,y,z}
\,\, &v_a  \hat G_B(t,t')  \hat v_a
\Big]
\hat R(-\theta)
\nonumber
\\
&\stackrel{!}{=}
\sum_{a=x,y,z}
\,\,\hat v_a \hat R(\theta) \hat G_B(t,t') \hat R(-\theta) \hat v_a,
\end{align}
which can be written as 
\begin{align}
I(\theta)
&\equiv
\sum_{a=x,y,z}
\!\!\!\!
\,\hat v_a(\theta) \hat G_B(t,t') \hat v_a(\theta)
\stackrel{!}{=}I(0),
\end{align}
with $\hat v_z(\theta)= \hat R(-\theta) \hat v_z \hat R(\theta)$. Using Eq.~\eqref{hoppx} and \eqref{hoppy}, we get
\begin{align}
\label{III}
I(\theta)=
\sum_{n=-2,0,2}
\!\!\!
\hat v_z(\theta+\frac{2n\pi}{3})  \hat G_B(t,t') \hat v_z(\theta+\frac{2n\pi}{3}).
\end{align}
Explicit evaluation gives
\begin{align}
\hat v_z(\phi)  \hat G \hat v_z(\phi)
&=
G_{11}
\begin{pmatrix}
s^4  
&
-s^3c 
\\
-s^3c 
&
s^2 c^2
\end{pmatrix}
+
G_{22}
\begin{pmatrix}
s^2 c^2  
&
-sc^3 
\\
-sc^3 
&
c^4
\end{pmatrix}\nonumber\\&
+
(G_{12}+G_{21})
\begin{pmatrix}
-s^3c 
&
-s^2c^2 
\\
-s^2c^2 
&
-sc^3
\end{pmatrix}
,
\label{Grot}
\end{align}
with $s\equiv \sin(\phi/2)$ and  $c\equiv \cos(\phi/2)$. 
After summation in Eq.~\eqref{III} we get,
\begin{align}
I(\theta)=
\frac{1}{8}
\begin{pmatrix}
9G_{11}+3G_{22} &
-3(G_{12}+G_{21})
\\
-3(G_{12}+G_{21})
&
3G_{11}+9G_{22}
\end{pmatrix},
\end{align}
independent of $\theta$. This implies that the DMFT solutions will be rotationally invariant in equilibrium, which is a consequence of the spacial mean-field character of the equations: In the real lattice, the mean-field pseudospin solution is rotationally invariant, and the rotational invariance is broken to the lattice point group only by the order-by-disorder mechanism, which takes into account fluctuations of the order parameter around the long range order (spin waves), which are not correctly captured in DMFT \cite{Nussinov2015}.







\begin{thebibliography}{99}
\bibitem{footnote01}
Other, slightly different parameterizations based on a non-rotationally invariant interaction Hamiltonian  have been considered in the literature [45]. 
\bibitem{footnote_mf}
Note that this is fundamentally different from photo-induced spin dynamics [48], where a change of exchange induces a spin precession only in the non-collinear ordered phase.
\end{thebibliography}

\end{document}